\documentclass[preprint,12pt]{elsarticle}

\usepackage{amssymb}
\usepackage{rotating}
\usepackage{listings}

\newcounter{bla}

\usepackage{xcolor}

\journal{Computer Physics Communications}

\begin{document}

\begin{frontmatter}



\title{{\bf FabSim: facilitating computational research through automation on large-scale and distributed e-infrastructures}}


\author[a,b]{Derek Groen\corref{author}}
\author[a]{Agastya Bhati}
\author[a]{James Suter}
\author[a]{James Hetherington}
\author[a]{Stefan Zasada}
\author[a]{Peter V. Coveney\corref{author}}

\cortext[author] {Corresponding authors.\\\textit{E-mail address:} Derek.Groen@brunel.ac.uk, p.v.coveney@ucl.ac.uk}
\address[a]{Centre for Computational Science, University College London, 20 Gordon street, London, WC1H 0AJ, United Kingdom}
\address[b]{Department of Computer Science, Brunel University, St John's Building, Kingston Lane, Uxbridge, UB8 3PH, United Kingdom}

\begin{abstract}
We present FabSim, a toolkit developed to simplify a range of computational
tasks for researchers in diverse disciplines. FabSim is flexible, adaptable, 
and allows users to perform a wide range of tasks with ease. It also
provides a systematic way to automate the use of resourcess, including HPC and distributed resources, 
and to make tasks easier to repeat by recording contextual information. To demonstrate this, 
we present three use cases where FabSim has enhanced our research productivity. 
These include simulating cerebrovascular bloodflow, modelling clay-polymer 
nanocomposites across multiple scales, and calculating ligand-protein binding affinities.
\end{abstract}





\begin{keyword}
automation; workflows; distributed computing; software; bloodflow modelling; molecular dynamics; lattice-Boltzmann; multiscale modelling; clay-polymer nanocomposites; ligand-protein binding

\end{keyword}

\end{frontmatter}



{\bf PROGRAM SUMMARY/NEW VERSION PROGRAM SUMMARY}

\begin{small}
\noindent
{\em Manuscript Title:} FabSim: facilitating computational research through automation on large-scale and distributed e-infrastructures                                      \\
{\em Authors:}                                                
  Derek Groen, Agastya Bhati, James Hetherington, James Suter, Stefan Zasada, Peter V. Coveney.\\
{\em Program Title:}                                          
  FabSim\\
{\em Journal Reference:}                                      \\
{\em Catalogue identifier:}                                   \\
{\em Licensing provisions:}                                   \\
  Lesser GNU Public License version 3.
{\em Programming language:}                                   
  Python.\\
{\em Computer:}                                               \\
  PC or Mac.
{\em Operating system:}                                       
  Unix, OSX.\\
{\em RAM:} 1 Gbytes                                              \\
{\em Supplementary material:}                                 \\
{\em Keywords:} 
automation, workflows, distributed computing, software, bloodflow modelling, molecular dynamics, lattice-Boltzmann, multiscale modelling, clay-polymer nanocomposites, ligand-protein binding.  \\
{\em Classification:} 3 Biology and Molecular Biology, 4 Computational Methods, 6.5 Software including Parallel Algorithms.\\
{\em External routines/libraries:} NumPy, SciPy, Fabric (1.5 or newer).  \\
{\em Subprograms used:}                                       \\
{\em Catalogue identifier of previous version:}*              \\
{\em Journal reference of previous version:}*                  \\
{\em Does the new version supersede the previous version?:}*   \\
{\em Nature of problem:}\\
Running advanced computations using remote resources is an activity that requires considerable time and human attention. 
These activities, such as organizing data, configuring software and setting up individual runs often vary slightly
each time they are performed. To lighten this burden, we required an approach that introduced little burden of its own to set up
and adapt, beyond which very substantial productivity ensues.  
   \\
{\em Solution method:}\\
We present a toolkit which helps to simplify and automate the activities which surround computational science research.
FabSim is aimed squarely at the experienced computational scientist, who can use the command line interface and a system
of modifiable content to quickly automate sets of research tasks.
   \\
{\em Reasons for the new version:}*\\
   \\
{\em Summary of revisions:}*\\
   \\
{\em Restrictions:}\\
FabSim relies on a command-line interface, and assumes some level of scripting knowledge from the user.
   \\
{\em Unusual features:}\\
FabSim has a proven track record of being easy to adapt. It has already been extensively adapted to facilitate
leading research in the modelling of bloodflow, nanomaterials, and ligand-protein binding.
   \\
{\em Additional comments:}\\
   \\
{\em Running time:}\\
FabSim can be used interactively, typically requiring a few seconds to perform a basic task.
   \\

\end{small}

\section{Introduction}

Research based on computational science and technology continues to advance 
at a rapid pace, driven in part by the continual evolution, and recent 
diversification, of computing infrastructures. At the top end, supercomputers 
are both highly parallel (at time of writing, the number one supercomputer has 3.1 million cores) 
and highly heterogeneous (four of the top ten supercomputers feature accelerators). 
Because computing infrastructures are growing in parallelism and becoming more diverse, 
we require sophisticated computational techniques to take full advantage of the power available. 

Recent advances in modelling methods, such as ensemble
and multiscale computing, allow us to solve complex problems more efficiently
using high-end infrastructures~\cite{Groen:2014,Wright:2014,Hoekstra:2014,Borgdorff:2014-2}.
These techniques tend to require users to construct, execute, validate, analyze 
and curate a number of different models
(and model executions) for each computation. Care must be taken when performing these
tasks, as a simple mistake can render the full computation useless. For example, 
computations may produce incorrect results, or produce too little information to allow 
for reproduction or replication.

In an era where compute resources are arguably easier to obtain than 
human resources, this requirement for continued human attention can become a 
bottleneck, limiting the pace of 
computational research to the number of person hours invested in it. Indeed, 
within our own research group we have realized that much of our daily work was
spent on the attention-requiring activities that accompany the need to
run ensemble or multiscale computations. To lighten this burden, we sought 
an approach that introduced little burden of its own to set up 
and adapt, beyond which very substantial productivity benefits ensue by
``automating away'' routine activities.

Here we present our approach, based on FabSim, for managing computations 
and automating the research tasks that accompany them. FabSim 
includes a software toolset, as well as a set of 
best practices which serve to aid researchers in maintaining a computational 
environment which is simple to use, navigate, interpret and modify. 
FabSim allows researchers to reduce the complexity of administrative tasks
to that of issuing single-line commands, and saves time by introducing
a systematic structure for curating input and output files, user and machine
configurations, as well as application execution instances. The tool is 
straightforward to use ``out of the box'', and as simple to customize. 


In Section~\ref{Sec:related}, we present a range of related research and
development activities. In Section~\ref{Sec:overview} we present FabSim, describe
its architecture as well as its key features for users and developers. In 
Section~\ref{Sec:usecases} we present three research activities where FabSim is
currently in use, and describe how FabSim is being deployed and adapted in these 
contexts. We provide concluding remarks in Section~\ref{Sec:conclusion}.

\section{Related work}\label{Sec:related}

To some extent, the functionality offered by FabSim is similar to that provided
by manifold grid middleware projects, developed over the last decade or more.
By allowing a computational scientist to run workloads on remote high
performance computing resources, FabSim shares functionality with middleware
toolkits such as Globus \cite{globus4}, Unicore \cite{unicore} and gLite
\cite{glite}. However, the focus of FabSim is very different to these
monolithic toolkits. Firstly, FabSim is aimed squarely at the experienced
computational scientist, presenting a command line interface that is easy for
the developer/scientist to extend. In this respect, FabSim shares similarities
with the Growl Toolkit \cite{GROWL}, an example of a lightweight middleware
system designed to address some of the shortcomings of other grid middlewares,
using both a combination of Web service components and wrapper scripts to
automate tasks performed by Globus grid middleware client tools. 

Additionally, FabSim is built on SSH, found on practically every Unix- or 
Linux-based system, and does not mandate the installation of an additional
heavyweight middleware stack on the resources being accessed. This means that
FabSim can be used widely on any HPC resource that supports SSH. 


FabSim is more directly comparable with tools such as Longbow~\cite{Longbow}, which is used 
for molecular dynamics pertaining to computational biology, and also aims to provide shorthand commands
to run applications (including ensembles) on distributed machines. Longbow is, however,
more limited in scope, as it provides no explicit support for multi-step workflows
or for tasks associated with code compilation and deployment.
Ruffus~\cite{Goodstadt:2010} is a light-weight Python tool which automates
complicated analysis activities, with less concern for computations on distributed
resources. Snakemake~\cite{Koester:2012} offers a workflow definition language,
and provides an execution environment. Snakemake workflows can be run remotely, 
although the tool does not provide further facilities for using distributed
resources (e.g., curating information for its users as to how to access different machines).


Many other tools provide functionalities that complement FabSim.  
Coupling environments such as 
MUSCLE 2~\cite{Borgdorff:2014}, DataSpaces~\cite{Docan:2012} and 
MPWide~\cite{Groen:2013-3} allow codes to efficiently exchange data at runtime, 
and can be used to speed up the remote execution of coupled tasks in FabSim.

In addition, there are a number of simulation environments which serve 
to combine functionalities from existing codes to construct and run multiscale 
simulations~\cite{Groen:2014}, such as AMUSE~\cite{Portegies:2013}, 
CouPe~\cite{CouPe}, MOOSE~\cite{Gaston:2009} and OASIS~\cite{Gregersen:2007}. 
In particular, the Application Hosting Environment~\cite{Zasada:2009} provides an easy-to-use
environment by centralizing the application deployment in the hands of one
expert-user, though it provides no automation for those users
who deploy new software. Both GEL~\cite{Lian:2005} and Swift~\cite{Wilde:2011} assist
in the coordination and efficient execution of large numbers of scripts, while
workflow engines such as Kepler~\cite{Taylor:2014,Ludascher:2006},
Taverna~\cite{Wolstencroft:2013} and GridSpace~\cite{Rycerz:2015} provide a
graphical environment to help users perform complex simulation workflows. 
Ludascher et al.~\cite{Ludascher:2009} provide a comprehensive
overview of key challenges and advances in workflow management and scientific
task automation.
A major strength of FabSim, compared to the aforementioned tools, is its 
strong focus on accelerating and simplifying development activities. Its aim is 
not only to simplify the execution of workflows that have been previously defined, 
but also to simplify the creation of new workflows, and indeed of new 
computational approaches in general.

\section{Overview of FabSim}\label{Sec:overview}

The central purpose of FabSim is to save time for computational researchers by
simplifying key tasks when performing computation-driven research. With FabSim we
achieve this by offering a set of useful functionalities in a highly
transparent and modifiable program structure. By pairing customizable software
with a set of best practices we provide researchers with a methodological
framework which helps them to perform a range of tasks in a simpler, quicker
and more systematic way.

\subsection{The software}

\begin{figure}[!t]
\centering
\includegraphics[width=5in]{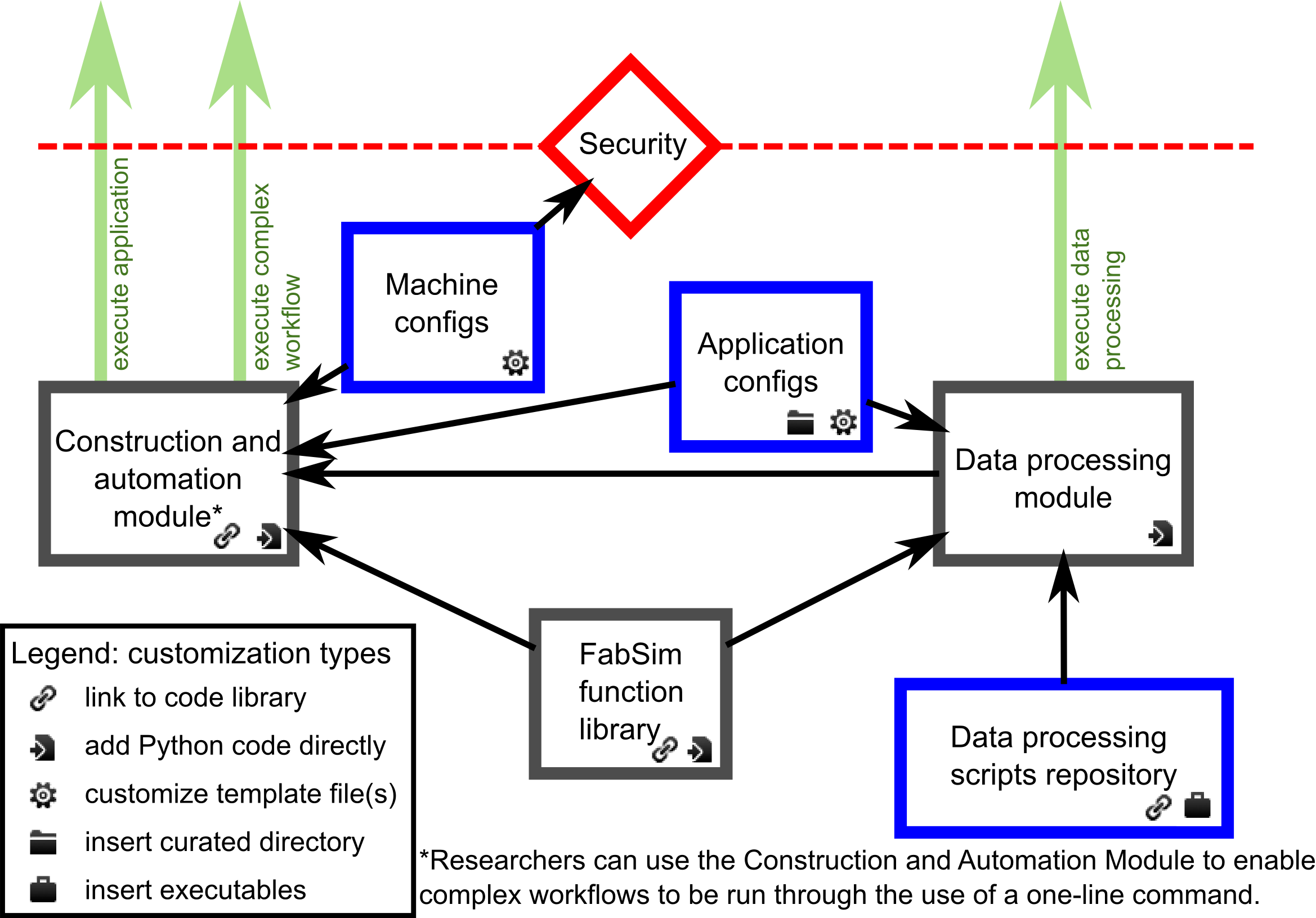}
\caption{Diagrammatic overview of the modules present in FabSim. Information dependencies are
indicated by black arrows (e.g., the data processing module makes use of information provided by the
application configs module). Security is managed using Paramiko (www.paramiko.org), which in turn
relies on SSH. All modules are installed on a user's local workstation. User activities involving remote
resources are indicated by light green arrows passing through the security layer (which is indicated 
by the red dotted line).  Small icons are provided by gentleface.com.}
\label{Fig:Architecture}
\end{figure}

FabSim is written in Python and requires no administrative privileges to
install. It relies on the Fabric~\cite{fabric} library to embed convenient,
light-weight and non-invasive mechanisms for accessing and managing remote
machines.  In addition, FabSim uses the YaML~\cite{yaml} library, which
provides features to work with a compact, intuitive and human-readable data
format. We provide an overview of the FabSim architecture in  Fig.~\ref{Fig:Architecture}.

\subsubsection{Using FabSim}

Once installed, all FabSim features are called from the terminal, using
commands that adhere to the following structure:
\\
\\ 
\texttt{fab <hostname>
<command>:<configuration\_name>,\\ 
<parameter1=x1>,(...),<parameterN=xn>} \\
\\
We describe a number of commands commonly used in FabSim in Table~\ref{Tab:FabCommands}.
Each function can be given parameters if convenient, or uses
machine-specific defaults if no parameters are specified by the user. We have
also defined a range of commands which are specific for FabSim in different domains. 
We discuss these commands separately in the examples provided in Section~\ref{Sec:usecases},
and describe how these functions can be used as part of more complex user-made 
commands.

\begin{sidewaystable}
\caption{List of commands commonly used in FabSim.}
\label{Tab:FabCommands}
\centering
\begin{tabular}{l|l}
\hline
command name & brief description\\
\hline
{\tt probe} & Probes target host for presence of a module with a given name \\
{\tt stat} & Provides a status report of jobs running on the target host \\
{\tt monitor} & Does a {\tt stat} on the target machine once every two minutes, and displays the output \\
{\tt fetch\_results} & Fetches all the application results from the target host and saves them on the local host \\
{\tt cancel} & Cancels a job on the target machine which in the queue or running \\
{\tt analyze} & Performs shallow data analysis of a results file \\
{\tt run\_job} & Runs an application job remotely \\
{\tt run\_ensemble} & Runs an ensemble of application jobs remotely (packed into one job if supported) \\
{\tt blackbox} & Runs a script through FabSim on the local machine \\
{\tt archive} & Archives a directory containing application results \\
\hline
{\tt cold} & Copies a source code to a remote resource, compiles and builds everything (application-specific)\\
{\tt <application name>} & Runs an application job using the \texttt{<application name>} code specifically (application-specific) 
\end{tabular}
\end{sidewaystable}

\subsubsection{Customizing FabSim}

A key strength of FabSim is its ease of customization. FabSim uses a
template/variable substitution system to enable users to easily introduce
customized scripts, and relies on the ease of use of Python to allow users to
define custom functionalities. The substitution system relies on a set of YaML
files, which specify the default values for all the important variables in
FabSim. Customized values can be assigned to these variables on a machine-specific
basis, on a user-specific basis, or both. When the user invokes a FabSim
command, the relevant variables are provided to the context of that command,
eliminating the need for users to specify these on the command line. 

FabSim also relies on a number of templates, for example to identify the formatting required
to make a correct header for a batch job script, or to flexibly insert commands
for executing a specific MPI implementation. When templates are used, FabSim 
uses the variables within its context to determine which values to insert into
the templates, and which templates to combine. For example, we can combine templates 
for executing specific applications with those for specific scheduling systems, 
without the need to define new templates for each combination of the two.

In Tab.~\ref{Tab:Components} we present the configuration and
software hooks that FabSim provides, each of which can be modified by the user.
Within FabSim we define a number of different scopes, as the features in FabSim
differ in their range of applicability. The base scope of FabSim is
general-purpose, which includes features that are deemed to be of use for any
user installing FabSim (e.g., job submission, file transfer and so on).  Features can be
added on this level by adding functions to fab.py. Machine-specific
customizations can be added in one of two separate scopes (``machine-specific''
and ``machine and user-specific''). Perhaps the scope most important to the
user is the domain-specific scope. The script files that reside in this scope
contain features that are specific to the user domain (e.g., multiscale
materials modelling, blood flow simulation, or molecular ligand-protein binding
calculations), but which can be reused for existing and future research problems.
In some projects we have been working on, the amount of customization in this scope has become so large that
the community adopted a modified name for the tool altogether (e.g.,
FabHemeLB and FabMD). Last, there is the scope which is problem-specific, which includes
bespoke scripts and features that are used for highly-specific research
purposes.

\begin{sidewaystable}
\label{Tab:Components}
\centering
\begin{tabular}{llll}
Name & Location & Scope & Summary of contents\\
\hline
config\_files & application configs & problem & input data for specific computations.\\
fab.py & constr. and auto. module & general & general Fabsim commands.\\
data\_proc/dataXX.py & data processing module & domain & data processing commands.\\
data\_proc/dataYY.py & data processing module & problem & data processing commands.\\
data\_proc/data.py & data processing module & general & data processing commands.\\
blackbox/ & data processing scripts repo & domain & plug-in scripts, binaries.\\
python/ & data processing scripts repo & domain & linkable Python plug-ins.\\
fabXX.py & function library & domain & FabSim commands.\\
fabXX/fabYY.py & function library & problem & FabSim commands.\\
machines.yml & machine configs & machine & default machine settings.\\
machines\_user.yml & machine configs & machine & user machine settings.\\
\hline
\end{tabular}
\caption{List of customizable components in FabSim. The name of the component is given in
the first column, the module where it is located (see 
Fig.~\ref{Fig:Architecture} for a diagrammatic overview of the modules) in the second column.
The scope within which the component is applied and customized is given in the third column. 
Possible scopes include general-purpose (general), domain-specific (domain), problem-specific (problem) and machine-specific (machine).
A short summary of the contents of the component is given in the fourth column.}
\end{sidewaystable}

\subsection{Best practices}

FabSim is accompanied with a number of best practices which aid the user in
maintaining a simple and consistent environment. Here we describe several
examples of best practices that are key to keeping FabSim simple to use and modify.


Machine-specific configurations, which are applicable to all users of that
machine, are defined in machines.yml, User-specific information for each
machine is stored locally in machines\_user.yml.

Custom-defined features for FabSim are variously general-purpose,
domain-specific, problem-specific or for single use. As best practice, we keep
general-purpose features in fab.py (or libraries that are included in fab.py), 
and domain-specific features in separate python files (e.g., fabNanoMD.py), 
distinct from any source files which contain problem-specific features.

\subsection{Remote execution}

FabSim allows users to perform computations on remote resources using one-line 
commands. For example, to launch an instance of the LAMMPS molecular dynamics
code on nodes on a supercomputer named ``exa'', one could use:

\begin{lstlisting}
fab exa lammps
\end{lstlisting}

As part of this command, FabSim stages in a directory with input files from
the local machine to the remote resource, and then uses SSH to submit the job
using the remote job scheduling system. One can use (or define) similar 
commands to perform tasks directly on the head node of the remote resource.
For example, to locate all modules containing the name ``lammps'' on the ``exa'' 
machine, one could use

\begin{lstlisting}
fab exa probe:lammps
\end{lstlisting}

It is also possible to combine the execution of multiple jobs on remote resources in a single
FabSim command (see Section~\ref{Sec:bac}), or to construct a chain of interdependent jobs interspersed with
local data processing tasks (see Section~\ref{Sec:nano}).

\subsection{Provenance, curation, reproducibility}

When a user performs a computation remotely with FabSim, a number of extra
actions are executed to help improve the repeatability and reproducibility of
the computation. First, FabSim stores the full internal context in YML format
in the results directory of the executed computation, allowing users to identify
FabSim variables that may have been wrongly set. Second, FabSim stores the environment
variables used at the remote resource in another text file, allowing users to
spot changes in the configuration of the remote resource between jobs. Third,
FabSim retains the generated job submission script for future reference. And
fourth, when FabSim creates results directories for submitted computations, it
allows users to modify the name of these directories using variables (e.g.,
code version number, time stamp, configuration file used, or the number of
cores used). In particular, by using time stamps in naming results directories
one can prevent new computations from overwriting results that were generated
by previous computations.

In our experience, we have frequently been able to repeat our runs using the
FabSim logging infrastructure. However, it may still occur that repeated 
computations lead to different results. We have experienced this occasionally when
existing modules residing on remote machines are recompiled with different 
settings or using a different compiler, or when the computations are repeated
using a different set of resources.

\subsection{Security}

Access to grid resources is usually secured through authentication and
authorization mechanisms based on X.509 certificates, a security credential
used to authenticate the user when accessing a grid. To access the resources on
a particular distributed e-infrastructure, the user needs a certificate
recognized by that infrastructure. Certificates are generally issued on a
national basis, by a national research certificate authority (CA). The
Interoperable Grid Trust Federation \cite{igtf} exists to ensure mutual trust
between different national certificate issuing bodies. This means that
certificates issued in one country will be accepted by e-infrastructure
resources based in a different country.

Efforts to address the usability of e-infrastructures have long been hampered by existing
security mechanisms imposed on users. Typically, these require a user to obtain
one or more digital certificates from a certificate authority, as well as to
maintain and renew these certificates as necessary. The difficulty in doing
this leads to widespread certificate sharing and misuse and a substantial
reduction in the number of potential users \cite{Beckles:2005}, and caused many HPC
resource providers and users to abandon grid middleware tools. This has in turn led to
secure certificate sharing mechanisms to be promoted which seek to ameliorate
some of the worst aspects of certificate misuse \cite{Zasada:2011}.

However, almost all HPC resources do support the SSH protocol to allow remote
users to access the machine. FabSim takes advantage of this, using the Fabric
library, to allow users to perform remote operations using basic SSH as the
transport middleware. As such, FabSim's security model is essentially the SSH
security model; public/private keys are used to authenticate FabSim operations on
target resources, and the \texttt{\textasciitilde/.ssh/known\_hosts} file is used to allow
users to configure mutual authentication.

Unlike grid based X.509 authentication, the setup is entirely controlled by the
end user. Typically on a grid using X.509 certificates, an administrator must
set up details of a user's certificate on their resources. SSH based
authentication requires that the user sets up their own keys on a target
resource. While key management does potentially increase the management
overhead of setting up FabSim security, the ubiquity of SSH mean that most
FabSim users will have already taken steps to set up SSH keys for the
resources that they wish to access.

The use of SSH keys and Fabric also means that FabSim can make use of Fabric
extensions for key management, such as the keymanager add-on \cite{keymanager},
which allows users to manage keys on multiple servers very easily, using Fabric
itself.

\section{Exemplar FabSim use cases}\label{Sec:usecases}

In this section we present exemplar FabSim use cases in three scientific
domains where FabSim has so far been applied. These include simulations of 
cerebrovascular bloodflow, multiscale simulations of clay-polymer nanocomposites,
and ensemble molecular dynamics simulations used to calculate ligand-protein binding affinities.

\subsection{Cerebrovascular bloodflow}\label{Sec:hemelb}


FabSim is widely used in combination with the HemeLB bloodflow
simulator~\cite{Mazzeo:2008,Groen:2013}, which we use to investigate the flow
properties of blood in arterial networks (e.g., a segment of the middle
cerebral artery or a network of vessels in the retina). HemeLB is
specifically optimized to efficiently model flow in sparse networks, and scales
to up to 49,152 cores~\cite{allinea,Groen:2013}.  Using both HemeLB and
FabSim, we have been able to make reliable predictions of the blood flow
properties in cerebral arterties~\cite{Itani:2015} and in mouse
retinas~\cite{Bernabeu:2014,Franco:2015}.
Accurate predictions of the blood flow under realistic conditions are essential
to gain a further understanding of important medical conditions, such as
aneurysm formation and tumour growth. For example, the risk of bleeding in the
brain appears to correlate with abnormal flow properties in the vicinity of
brain aneurysms, such as a high wall shear stress~\cite{Cebral:2011}.

\begin{figure}[!t]
\centering
\includegraphics[width=5in]{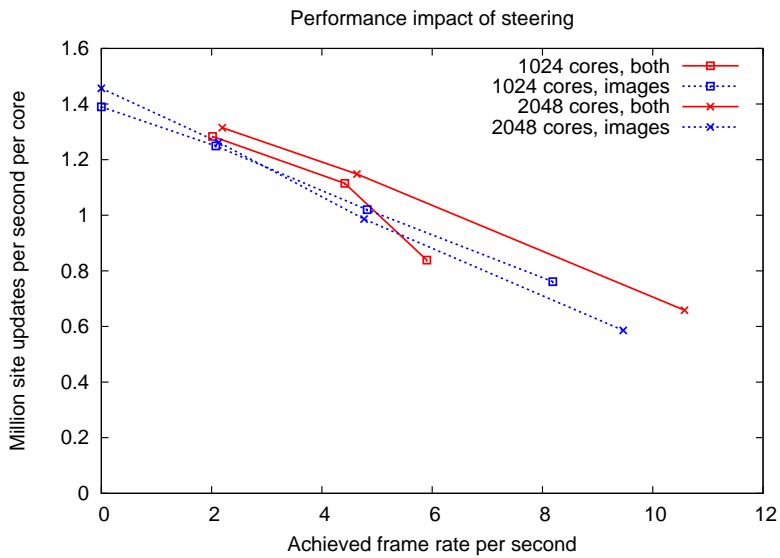}

\caption{Example of a benchmark which was systematically acquired using
FabSim on the HECToR supercomputer at EPCC in Edinburgh, United Kingdom. We
show the performance impact of running HemeLB with a connected steering client~\cite{Mazzeo:2008}.
We show results for 1024 and 2048 cores without steering client (plotted at
frame-rate zero), with the client used only for image streaming (images, dotted
lines) and with the client used both for image streaming and steering the
HemeLB simulation (both, solid lines).  This figure is reproduced from Groen et
al.~\cite{Groen:2013}, in which the significance of this data is discussed in
detail.}

\label{Fig:hemelb-example-result}
\end{figure}

Among other things, we use FabSim to help make HemeLB easier to install and use for
inexperienced researchers. HemeLB is in use across a range of supercomputer
platforms, and many of its applications require complex workflows consisting of
multiple simulations. Here, the HemeLB-extended version of FabSim allows users
to systematically run, curate and analyze sets of simulations (see
Fig.~\ref{Fig:hemelb-example-result} for an example). This makes it
considerably easier to perform scalability and accuracy studies which require
a large number of simulation runs~\cite{Groen:2013,Nash:2014}. 

\begin{figure}[!t]
\centering
\includegraphics[width=5in]{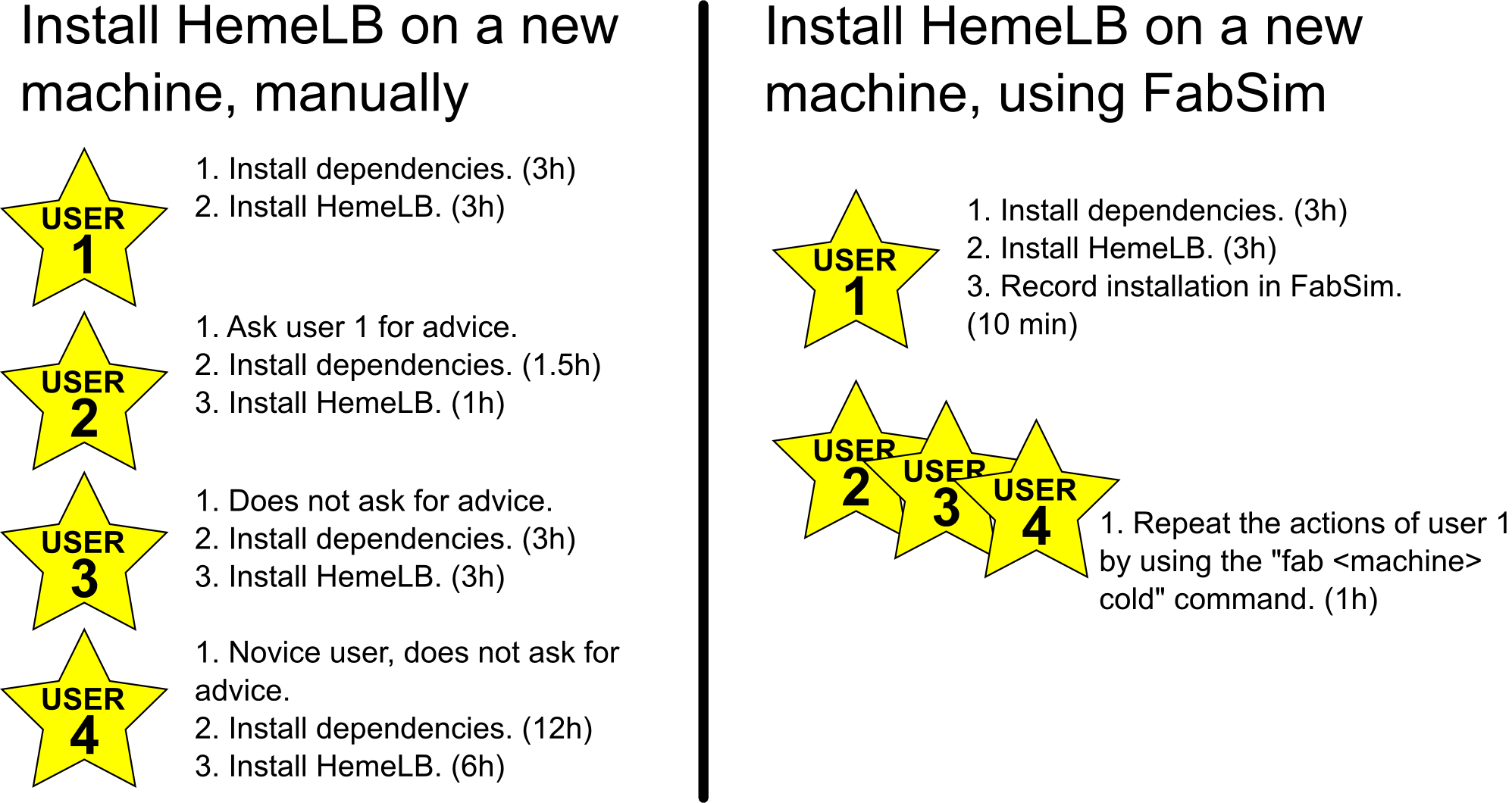}
\caption{Example of the speed-up obtained through the use of FabSim. We present typical tasks performed by a team of researchers when installing HemeLB manually (on the left), or using FabSim (right).}
\label{Fig:hemelb-install}
\end{figure}

In addition, for HemeLB we use the FabSim templating system to automate the
installation of HemeLB on new supercomputers. This process was previously
labour-intensive as the existing CMake system only permits a limited amount of
automation, and lacks an intuitive way to store machine-specific information
about the required configuration flags and environment settings. We present the 
added value of FabSim in the remote installation of
HemeLB in Fig.~\ref{Fig:hemelb-install}. Here, FabSim saves time by
allowing researchers to provision installation details in a compact and
readable way, thereby automating the installation process for future users. 

\begin{figure}[!t]
\centering
\includegraphics[width=5in]{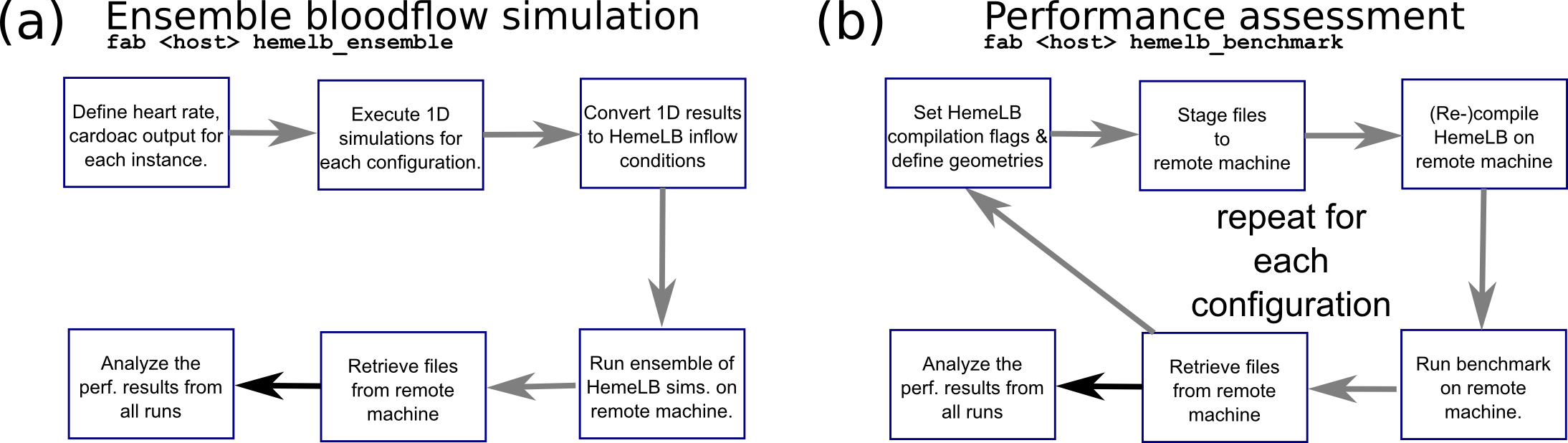}
\caption{Workflows which have been automated using FabHemeLB. (a) Diagram of the workflow used to perform ensemble multiscale simulations of blood flow in middle cerebral arteries (presented in detail in~\cite{Itani:2015}). (b) A commonly applied workflow to do systematic performance tests (applied for example in Groen et al.~\cite{Groen:2014}).}
\label{Fig:hemelb-example}
\end{figure}

FabHemeLB also provides a range of domain-specific features, primarily to
enable automated execution of ensemble multiscale simulations. In
Fig.~\ref{Fig:hemelb-example} we present two example workflows which are
supported by FabHemeLB. In the workflow on the left we first collect
patient-specific heart parameters (e.g., heart rate, cardiac output volume) at
different levels of physical activity. We then use a 1D model (The Python
Network Solver~\cite{Manini:2014}) to calculate the expected inflow profile in a
cerebral artery, which is used with HemeLB to model a patient-specific
cerebral arterial network in 3D. We then collect and analyze the results of our
ensemble of simulations using a one-line FabSim command. Through adopting
FabSim, we enabled a new kind of simulation analysis with HemeLB, in which we
can quickly run a set of simulations on a large supercomputer to compare wall
shear stresses in arterial bloodflow as a function of exercise intensity. We
present a detailed description of this workflow, used on the 2.6 PFLOPs ARCHER
supercomputer in Edinburgh, UK, in Itani et al.~\cite{Itani:2015}. 


\subsection{Multiscale modelling of clay-polymer nanocomposite materials}\label{Sec:nano}

Multiscale modelling approaches offer large advantages in the domain of 
materials modelling. Here, a key challenge in the field is to predict large scale 
materials properties, whilst taking into
account the chemical specificity of its constituent atoms and molecules,
as well as processing conditions. 

We have created an adapted version of FabSim (named FabMD) to construct and
apply a multiscale modelling methodology for the study of clay-polymer
nanocomposite materials~\cite{Suter:2015,Suter:2015-2}. These composite materials, due to a
combination of their low density with superior materials properties, have
already been applied in industries such as packaging, automotive, aviation, and drug
transport~\cite{Ray:2014}. However, it is costly, time-consuming and labour-intensive 
to search for new materials using experimental approaches, whereas
multiscale simulations are relatively fast and cheap. As a result, using
multiscale approaches we can identify systems that are likely to possess
superior materials properties, focussing the laboratory searches for much greater
efficiency and hence accelerated discovery.

\begin{figure}[!t]
\centering
\includegraphics[width=5in]{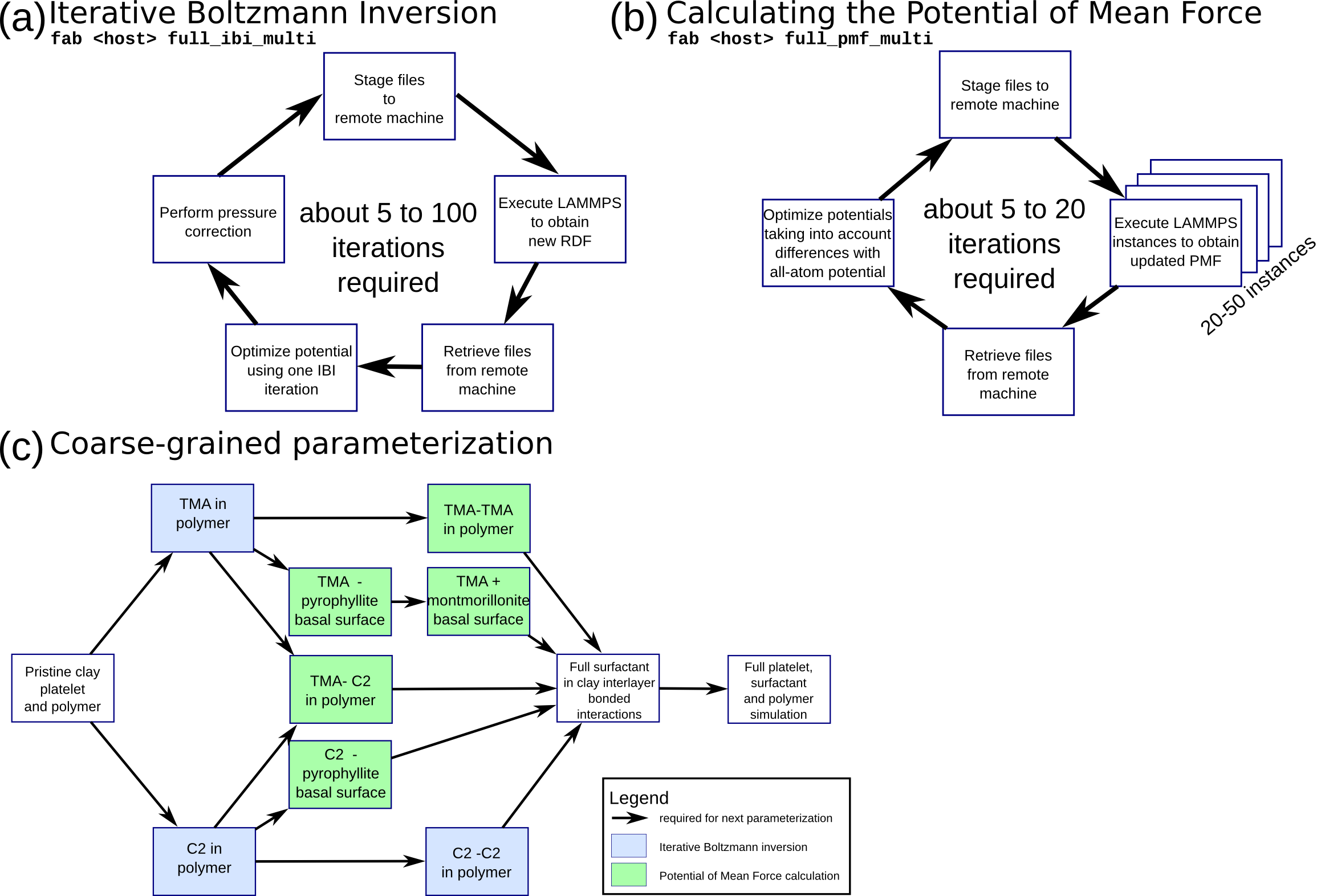}
\caption{Workflows which have been automated using FabMD to create a coarse-grained model of a mixture of montmorillonite clay and polymers. (a) Workflow used to perform Iterative Boltzmann Inversions. (b) Workflow used to calculate the Potential of Mean Force. (c) Diagrammatic overview of the steps involved to do a full coarse-grained parameterization (applied e.g. by Suter et al.~\cite{Suter:2015}). Here the molecule names have been abbreviated: TMA for tetra-methyl ammonium and C2 for ethane particles.}
\label{Fig:fabnano}
\end{figure}

We present a diagrammatic description of our multiscale materials modelling workflows in
Figure~\ref{Fig:fabnano}. We use FabSim to coarse-grain our clay-polymer
systems, and rely on two techniques to determine accurate course-grained potential 
parameterizations: Iterative Boltzmann Inversion (IBI) and
Potential of Mean Force (PMF) calculations. In the IBI procedure we
iteratively adjust pair potentials, and launch a single simulation per iteration 
to determine the radial distribution function which results from using these potentials, until they match the distribution functions from all-atom molecular dynamics simulations~\cite{Suter:2015} within a desired tolerance. 
IBI is computationally efficient, but can be inaccurate, particularly when the
number of particles is small (e.g., solutes dissolved in solution) or when the interaction potentials are
particularly attractive.  Whenever IBI is ineffective, we instead apply PMF, which
is computationally more expensive but more robust. When using PMF, each
simulation has its particles constrained at a given distance; this is done to
calculate the mean force between two particle types at a given distance. For
each PMF iteration, we then perform a set of $\sim$20-40 simulations with the particles constrained at various distances. The large
number of relatively small scale iterative computations required for these 
coarse-grained potential parameterizations lends itself well to automation using FabSim.

\begin{figure}[!t] \centering
\includegraphics[width=5in]{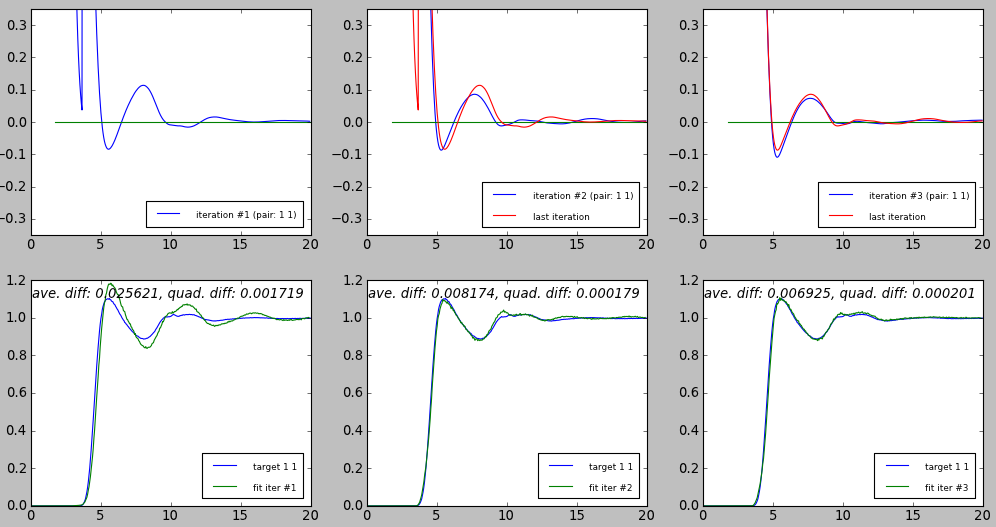} 

\caption{Example output of the first four Iterative Boltzmann Inversions (IBI),
used to parameterize polymer-polymer interactions.  The changes in
potential obtained through the procedure are shown in the top row for
iterations 1 to 3 (left to right). The resulting changes in the radial
distribution functions (RDF) for each potential are provided in the bottom row
for iterations 1 to 3 (left to right).  Here we also plot the desired
RDF, which we obtained from an atomistic simulation of the
same system (``target''), as well as two error measures (the mean absolute
difference and the mean squared difference) which can be used as convergence
criteria. For reasons of simplicity we have omitted information on the pressure
correction, which is simultaneously performed in these IBI
iterations~\cite{Suter:2015}.}

\label{Fig:fabnano-example-result} 
\end{figure}

As an example, we summarize the input and output of several IBI iterations in 
Figure~\ref{Fig:fabnano-example-result}. Here we performed four iterations to 
parameterize a suitable coarse-grained potential between two polymer 
particles~\cite{Suter:2015}. We provide the potentials which were generated using IBI for 
each of the simulation runs in the upper row of plots, and the radial 
distribution functions (RDF) that resulted from the simulation runs, and which were
used to generate the potentials for the next IBI iteration, in the bottom row.
We are also able to calculate the error between the obtained and desired RDF (the 
latter comes from all-atom molecular dynamics simulations~\cite{Suter:2015},
and use this as a convergence criterion. In addition to minimizing this error,
we also perform the same procedure to ensure the system pressure becomes close to
1 atmosphere (plots of the pressure fluctuations have been omitted to preserve 
space). FabMD allowed us to perform these procedures in a quick, automated 
fashion, providing adaptable quick-hand commands to generate 
suitable potentials for new systems as our study proceeded.

\subsection{Calculating ligand-protein binding affinities from ensemble molecular dynamics}\label{Sec:bac}

The ability to calculate the free energy of binding a lead compound with a target
protein, also known as the binding affinity, is of great importance in the field of
personalised medicine and drug discovery as, in most cases, it forms the basis
of ranking drugs/lead molecules based on their potency. Several \textit{in
silico} methods are available to calculate binding affinities, but the field has
gained a degree of notoriety since, frequently, results reported in the
literature have not been repeatable by others~\cite{Wright:2014,Wan:2015}. This lack
of reproducibility is due to the insufficient sampling of phase space, arising
from the extreme sensitivities of calculated properties to the initial
conditions of a molecular dynamics system. Such unreproducible binding
affinities are clearly unreliable for medical or industrial applications.

Using an ensemble simulation approach we have shown that, if the entire protocol of
binding affinity calculation is repeated a sufficient number of times for a
biomolecular system, then the computed binding affinities have a Gaussian
frequency distribution and their ensemble average is the theoretical estimate of the binding
affinity for that biomolecular system, with bootstrapping providing tight error bounds. 
Such precise and reproducible binding affinity estimates are expected to be
useful in drug discovery and drug selection in personalized medicine. 

\begin{figure}[!t]
\centering
\includegraphics[width=5in]{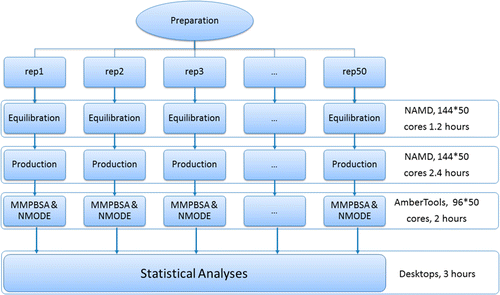}
\caption{
Example of an ESMACS workflow, automated using FabSim, as it is employed in the pMHC binding affinity
prediction, reproduced from Wan et al.~\cite{Wan:2015}. Here, 50 trajectory
calculations (known as replicas) are performed concurrently in three phases
(equilibration, production, MMPBSA\&NMODE), each replica using up to 144 cores.
After these calculations have completed, the peptide-protein binding affinity
is obtained through statistical analysis.  We show the ensemble simulations
required in a single trajectory calculation for which we need to have access to
7,200 cores for simulations and to 4,800 cores for free energy calculations.
For a concurrent three-trajectory case we need 19,200 cores for simulations and
14,400 cores for free energy calculations. The workflow requires approximately
9 hours to complete, given that sufficient resources are available.  To compute
more than one binding affinity concurrently, one needs to multiply the
requirement by the number of peptides of interest.} 
\label{Fig:fabbio1}
\end{figure}

Our ensemble simulation approach to control errors and ensure repoducibility,
which uses the Binding Affinity Calculator (BAC~\cite{Sadiq:2008-2,Sadiq:2010,Wright:2014}) in combination
with FabSim, is shown in Figure~\ref{Fig:fabbio1}.  In this approach we launch
multiple instances of the given molecular model, each of which is called a
``replica''.  Each replica uses initial atomic velocities which are randomly
drawn from a Maxwell-Boltzmann probability distribution.  For each replica we
then perform equilibration, production molecular dynamics, followed by
post-processing of the MD trajectories to determine the free energy of binding
based on our ESMACS protocol~\cite{Wan:2015}. ESMACS uses MMPBSA and NMODE
algorithms available in AMBER~\cite{Pearlman:1995} as well as the free energy of
association and, if required, an adaptation energy~\cite{Wan:2015}. As a last step, we
perform a statistical analysis on the collected results to report binding
affinities with error estimates. Thus, the methodology requires executing a
workflow comprising numerous steps on distributed (remote) machines, a process
which we have automated using an adapted version of FabSim (named FabBioMD). If
performed manually, this activity would have a very substantial manual overhead, 
reducing the rate of progress and being prone to human errors. On a
sizeable HPC resource, the approach has the considerable advantage that all
replicas can be run concurrently: in the time it takes to run one, we can run
all of them.

FabBioMD includes specific functions to perform production molecular
dynamics simulations and data analytics in the form of free energy calculations. 
These tasks can be performed on distributed (remote) resources
as needed.  All the input parameters have their default values
defined, but users are also allowed to set custom values using command
line parameters. We already used our approach on a number of supercomputers, including
the ARCHER supercomputer at EPCC in Edinburgh, UK and the BlueWonder supercomputer at
STFC in Daresbury, UK. Little effort is required to install it on
other supercomputers, as FabBioMD uses job-specific
and machine-specific templates of remote job submission scripts with variables
embedded at appropriate places. Users are able to modify these
templates to cater for their specific needs. This makes FabBioMD highly flexible, 
user friendly software. Moreover, the installation and usage of
FabBioMD is kept very simple, ensuring that non-speciliast users can easily simplify 
their computational activities. FabBioMD currently supports NAMD for
ensemble MD simulation and AMBER for the calculation of free energy contributions to the 
binding affinity based on MMPBSA~\cite{Kollman:2000} and normal mode methods respectively.

A number of related tools introduce automated workflows into free energy
calculations.  These tools include for example the Amber
suite~\cite{Case:2005}, FESetup~\cite{FESetup}, free energy workflow
(FEW)~\cite{Homeyer:2013}, free energy perturbation (FEP)
workflow~\cite{Wang:2015}, as well as an automation approach using
Copernicus~\cite{Pronk:2015} in combination with Gromacs~\cite{Pronk:2013}.  In
general, these tools are limited to the traditional one-off MD simulation
approach, which is limited by unreliable free energy predictions, and do not
support the ensemble simulation approach we described here.

\begin{figure}[!t]
\centering
\includegraphics[width=5in]{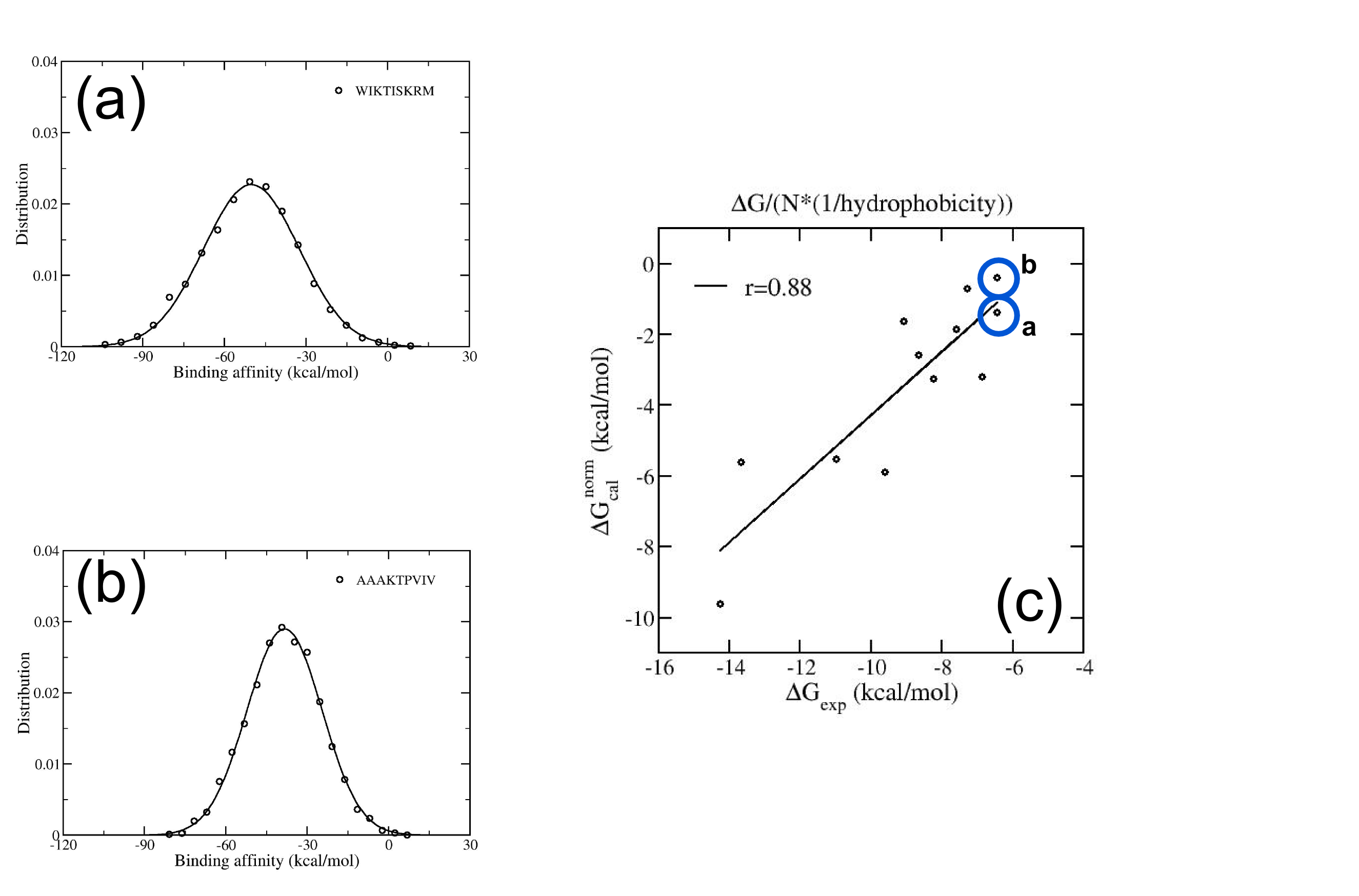}
\caption{a and b) Example outcomes of two peptide binding affinity calculations
(WIKTISKRM and AAAKTPVIV) using ESMACS within BAC (from Wan et al.~\cite{Wan:2015}). Each simulation
performed resulted in one data point; 50 simulations were run in total leading
to the Gaussian frequency distribution shown. c): Comparison of the peptide
binding affinity calculation results with experimental
results. Each data point consists of one binding affinity calculation, with the
two points in the top right corresponding respectively, 
from top to bottom, to the outcomes shown in (b) and (a). These results are discussed in detail in Wan et
al.~\cite{Wan:2015}.}
\label{Fig:fabbio2}
\end{figure}

We recently published a set of predictions of peptide-MHC (MHC is the 
Major Histocompatibiliy Complex protein) binding affinities using this ensemble
simulation approach~\cite{Wan:2015}, where FabBioMD was employed to
automate the process. The binding between a peptide and MHC is central to the
human immune response. In this study, a set of 12 different peptide sequences
bound to \texttt{HLA-A*02:01 MHC} allele were selected, and their calculated binding
affinities were compared with those determined experimentally. The size of the
ensemble was taken to be 50. The free energy was calculated using the ESMACS protocol, via the BAC workflow tool.
In Fig.~\ref{Fig:fabbio2}a and b we present an example normalised
frequency distribution of the ensemble of binding affinities of two of the 12
peptides. Each distribution is a Gaussian and corresponds to a single point in
Fig.~\ref{Fig:fabbio2}c, where we provide the correlation between
experimental binding affiinities and predictions using ESMACS (normalised by the number of heavy atoms and the peptide's cumulative
hydrophobicity for all 12 peptides). 

\section{Conclusion}\label{Sec:conclusion}

In the present work, we have described and made available FabSim, a new
approach to reduce the complexity of tasks associated with computational
research. We illustrate the scope and flexibility of FabSim by presenting its
application in three diverse domains where it has been, and continues to be,
used to simplify computational tasks and to improve their reproducibility. Our
use cases in bloodflow modelling, materials modelling and binding affinity
calculation provide evidence that FabSim benefits computational research on a
generic level.

In the area of brain bloodflow, we have described how FabSim can be used to do
systematic benchmarking, to execute an ensemble of multiscale simulations, and
to simplify the deployment of HemeLB on remote machines. In the nanomaterials
area, we have shown how FabSim automates iterative paramerization of
coarse-grained potentials, and allows us to systematically model the
self-assembly of layered composite materials with chemical specificity. FabSim
has also been applied to streamline the calculation of ligand-protein binding affinities
through our Binding Affinity Calculator,
allowing users to automatically launch ensemble computations, and thereby 
controlling uncertainties and producing reproducible results.
In all cases, FabSim assists in the curation of run data by furnishing information
about the job specification and the environment variables.
The extent to which FabSim has been applied and adapted in these three domains serves
to demonstrate its flexibility and ease of adoption. Indeed, using FabSim we
have been able to publish our research findings in leading scientific journals in each domain. 

As part of this software paper, we provide the FabSim code base to the scientific
community under a LGPL version 3 licence. As part of so doing, we hope to encourage
computational researchers to begin ``automating away'' some of their more
tedious tasks, and to free up more human effort for advancing science.





\section*{Acknowledgements}

We thank Dr Shunzhou Wan for his help in constructing the Binding Affinity
Calculation section of this article, and Miguel Bernabeu, Rupert Nash,
Sebastian Schmieschek, Mohamed Itani and Hywel Carver for their contributions
to FabHemeLB. This work was funded in part by the EU FP7 MAPPER, CRESTA,
P-medicine and VPH-SHARE project (grant numbers 261507, 287703, 270089,
269978), by the EU H2020 ComPat project (grant no. 671564) by EPSRC via the
2020 Science Programme (EP/I017909/1), the Qatar National Research Fund (grant
number 09–260–1–048), MRC Bioinformatics project (MR/L016311/1) and the UCL
Provost. AB is funded by the INLAKS Foundation Scholarship and a UCL Overseas
Research Studentship Award (2014-2017). 
Supercomputing time was provided by the Hartree Centre (Daresbury Laboratory) on BlueJoule and BlueWonder via the CGCLAY project, and on HECToR and ARCHER, the UK national supercomputing facility at the University of Edinburgh, via EPSRC through grants EP/F00521/1, EP/E045111/1, EP/I017763/1 and the UK Consortium on Mesoscopic Engineering Sciences (EP/L00030X/1).

\bibliographystyle{elsarticle-num}
\bibliography{Library}

\begin{thebibliography}{10}
\expandafter\ifx\csname url\endcsname\relax
  \def\url#1{\texttt{#1}}\fi
\expandafter\ifx\csname urlprefix\endcsname\relax\def\urlprefix{URL }\fi
\expandafter\ifx\csname href\endcsname\relax
  \def\href#1#2{#2} \def\path#1{#1}\fi

\bibitem{Groen:2014}
D.~Groen, S.~J. Zasada, P.~V. Coveney, Survey of multiscale and multiphysics
  applications and communities, {IEEE} Comput. Sci. Eng. 16~(2) (2014) 34--43.

\bibitem{Wright:2014}
D.~W. Wright, B.~A. Hall, O.~A. Kenway, S.~Jha, P.~V. Coveney, Computing
  clinically relevant binding free energies of {HIV-1} protease inhibitors,
  Journal of Chemical Theory and Computation 10~(3) (2014) 1228--1241.

\bibitem{Hoekstra:2014}
A.~Hoekstra, B.~Chopard, P.~V. Coveney, Multiscale modelling and simulation: a
  position paper, Philosophical Transactions of the Royal Society A:
  Mathematical, Physical and Engineering Sciences 372~(2021).

\bibitem{Borgdorff:2014-2}
J.~Borgdorff, M.~Ben~Belgacem, C.~Bona-Casas, L.~Fazendeiro, D.~Groen,
  O.~Hoenen, A.~Mizeranschi, J.~L. Suter, D.~Coster, P.~V. Coveney,
  W.~Dubitzky, A.~G. Hoekstra, P.~Strand, B.~Chopard, Performance of
  distributed multiscale simulations, Philosophical Transactions of the Royal
  Society A: Mathematical, Physical and Engineering Sciences 372~(2021).

\bibitem{globus4}
I.~Foster, {Globus Toolkit Version 4: Software for Service-Oriented Systems},
  Journal of Computer Science and Technology 21~(4) (2006) 513--520.

\bibitem{unicore}
{The UNICORE Project}, http://www.unicore.org.

\bibitem{glite}
{gLite Middleware}, http://glite.web.cern.ch/glite/.

\bibitem{GROWL}
M.~Hayes, L.~Morris, R.~Crouchley, D.~Grose, T.~Van~Ark, R.~Allan, J.~Kewley,
  Growl: A lightweight grid services toolkit and applications, in: 4th UK
  e-Science All Hands Meeting,” Nottingham, UK, 2005.

\bibitem{Longbow}
{Longbow - http://www.hecbiosim.ac.uk/wikis/index.php/Longbow} (2015).

\bibitem{Goodstadt:2010}
L.~Goodstadt, Ruffus: a lightweight python library for computational pipelines,
  Bioinformatics 26~(21) (2010) 2778--2779.

\bibitem{Koester:2012}
J.~Koester, S.~Rahmann, Snakemake -- a scalable bioinformatics workflow engine,
  Bioinformatics 28~(19) (2012) 2520--2522.

\bibitem{Borgdorff:2014}
J.~Borgdorff, M.~Mamonski, B.~Bosak, K.~Kurowski, M.~Ben~Belgacem, B.~Chopard,
  D.~Groen, P.~V. Coveney, A.~G. Hoekstra, Distributed multiscale computing
  with {MUSCLE} 2, the multiscale coupling library and environment, Journal of
  Computational Science 5~(5) (2014) 719--731.

\bibitem{Docan:2012}
C.~Docan, M.~Parashar, S.~Klasky, Dataspaces: an interaction and coordination
  framework for coupled simulation workflows, Cluster Computing 15~(2) (2012)
  163--181.

\bibitem{Groen:2013-3}
D.~Groen, S.~Rieder, S.~P. Zwart, {MPWide}: a light-weight library for
  efficient message passing over wide area networks, Journal of Open Research
  Software 1~(1) (2013) e9.

\bibitem{Portegies:2013}
S.~F. Portegies~Zwart, S.~L. McMillan, A.~van Elteren, F.~I. Pelupessy,
  N.~de~Vries, Multi-physics simulations using a hierarchical interchangeable
  software interface, Computer Physics Communications 184~(3) (2013) 456--468.

\bibitem{CouPe}
{CouPE - sites.google.com/site/coupempf/} (2015).

\bibitem{Gaston:2009}
D.~Gaston, C.~Newman, G.~Hansen, D.~Lebrun-Grandié, Moose: A parallel
  computational framework for coupled systems of nonlinear equations, Nuclear
  Engineering and Design 239~(10) (2009) 1768 -- 1778.

\bibitem{Gregersen:2007}
J.~B. Gregersen, P.~J.~A. Gijsbers, S.~J.~P. Westen, {OpenMI: Open modelling
  interface}, Journal of Hydroinformatics 9~(3) (2007) 175--191.

\bibitem{Zasada:2009}
S.~J. Zasada, P.~V. Coveney, Virtualizing access to scientific applications
  with the application hosting environment, Computer Physics Communications
  180~(12) (2009) 2513 -- 2525.

\bibitem{Lian:2005}
C.~C. Lian, F.~Tang, P.~Issac, A.~Krishnan, {GEL}: Grid execution language,
  Journal of Parallel and Distributed Computing 65~(7) (2005) 857--869.

\bibitem{Wilde:2011}
M.~Wilde, M.~Hategan, J.~M. Wozniak, B.~Clifford, D.~S. Katz, I.~Foster, Swift:
  {A} language for distributed parallel scripting, Parallel Computing 39~(9)
  (2011) 633--652.
\newblock \href {http://dx.doi.org/10.1016/j.parco.2011.05.005}
  {\path{doi:10.1016/j.parco.2011.05.005}}.

\bibitem{Taylor:2014}
I.~J. Taylor, E.~Deelman, D.~B. Gannon, M.~Shields, Workflows for e-Science:
  scientific workflows for grids, Springer Publishing Company, Incorporated,
  2014.

\bibitem{Ludascher:2006}
B.~Lud{\"a}scher, I.~Altintas, C.~Berkley, D.~Higgins, E.~Jaeger, M.~B. Jones,
  E.~A. Lee, J.~Tao, Y.~Zhao, Scientific workflow management and the kepler
  system, Concurrency and Computation: Practice and Experience 18~(10) (2006)
  1039--1065.

\bibitem{Wolstencroft:2013}
K.~Wolstencroft, R.~Haines, D.~Fellows, A.~Williams, D.~Withers, S.~Owen,
  S.~Soiland-Reyes, I.~Dunlop, A.~Nenadic, P.~Fisher, et~al., The {T}averna
  workflow suite: designing and executing workflows of web services on the
  desktop, web or in the cloud, Nucleic Acids Research (2013) 328.

\bibitem{Rycerz:2015}
K.~Rycerz, M.~Bubak, E.~Ciepiela, D.~HareÅ¼lak, T.~GubaÅ‚a, J.~Meizner,
  M.~Pawlik, B.~Wilk, Composing, execution and sharing of multiscale
  applications, Future Generation Computer Systems 53~(0) (2015) 77 -- 87.

\bibitem{Ludascher:2009}
B.~Lud{\"a}scher, I.~Altintas, S.~Bowers, J.~Cummings, T.~Critchlow,
  E.~Deelman, D.~D. Roure, J.~Freire, C.~Goble, M.~Jones, S.~Klasky,
  T.~McPhillips, N.~Podhorszki, C.~Silva, I.~Taylor, M.~Vouk, Scientific
  process automation and workflow management, Scientific Data Management:
  Challenges, Existing Technology, and Deployment, Computational Science Series
  230 (2009) 476--508.

\bibitem{fabric}
J.~Forcier, Fabric; http://www.fabfile.org (2014).

\bibitem{yaml}
{YAML - www.yaml.org} (2015).

\bibitem{igtf}
{IGTF: Interoperable Global Trust Federation}, https://www.igtf.net/.

\bibitem{Beckles:2005}
B.~Beckles, V.~Welch, J.~Basney, Mechanisms for increasing the usability of
  grid security, International Journal of Human-Computer Studies 63~(1/2)
  (2005) 74--101.

\bibitem{Zasada:2011}
S.~J. Zasada, A.~N. Haidar, P.~V. Coveney, On the usability of grid middleware
  and security mechanisms, Philosophical Transactions of the Royal Society A:
  Mathematical, Physical and Engineering Sciences 369~(1949) (2011) 3413--3428.

\bibitem{keymanager}
{Fabric SSH Keymanager}, https://github.com/farridav/keymanager.

\bibitem{Mazzeo:2008}
M.~D. Mazzeo, P.~V. Coveney, {HemeLB: A high performance parallel
  lattice-{B}oltzmann code for large scale fluid flow in complex geometries},
  Computer Physics Communications 178~(12) (2008) 894--914.

\bibitem{Groen:2013}
D.~Groen, J.~Hetherington, H.~B. Carver, R.~W. Nash, M.~O. Bernabeu, P.~V.
  Coveney, Analyzing and modeling the performance of the {HemeLB}
  lattice-{B}oltzmann simulation environment, Journal of Computational Science
  4~(5) (2013) 412 -- 422.

\bibitem{allinea}
{Neurological simulation milestone reached after UCL embraces Allinea’s tools
  on UK’s largest supercomputer},
  http://www.allinea.com/news/201406/neurological-simulation-milestone-reached-after-ucl-embraces-allinea\%E2\%80\%99s-tools-uk\%E2\%80\%99s.

\bibitem{Itani:2015}
M.~A. Itani, U.~D. Schiller, S.~Schmieschek, J.~Hetherington, M.~O. Bernabeu,
  H.~Chandrashekar, F.~Robertson, P.~V. Coveney, D.~Groen, An automated
  multiscale ensemble simulation approach for vascular blood flow, Journal of
  Computational Science 9 (2015) 150--155.

\bibitem{Bernabeu:2014}
M.~O. Bernabeu, M.~L. Jones, J.~H. Nielsen, T.~Kr{\"u}ger, R.~W. Nash,
  D.~Groen, S.~Schmieschek, J.~Hetherington, H.~Gerhardt, C.~A. Franco, P.~V.
  Coveney, Computer simulations reveal complex distribution of haemodynamic
  forces in a mouse retina model of angiogenesis, Journal of The Royal Society
  Interface 11~(99).

\bibitem{Franco:2015}
C.~A. Franco, M.~L. Jones, M.~O. Bernabeu, I.~Geudens, T.~Mathivet, A.~Rosa,
  F.~M. Lopes, A.~P. Lima, A.~Ragab, R.~T. Collins, L.-K. Phng, P.~V. Coveney,
  H.~Gerhardt, Dynamic endothelial cell rearrangements drive developmental
  vessel regression, PLoS Biol 13~(4) (2015) e1002125.

\bibitem{Cebral:2011}
J.~R. Cebral, F.~Mut, J.~Weir, C.~Putman, Quantitative characterization of the
  hemodynamic environment in ruptured and unruptured brain aneurysms, American
  Journal of Neuroradiology 32~(1) (2011) 145--151.

\bibitem{Nash:2014}
R.~W. Nash, H.~B. Carver, M.~O. Bernabeu, J.~Hetherington, D.~Groen,
  T.~Kr\"{u}ger, P.~V. Coveney, Choice of boundary condition for
  lattice-boltzmann simulation of moderate-reynolds-number flow in complex
  domains, Phys. Rev. E 89 (2014) 023303.

\bibitem{Manini:2014}
S.~Manini, L.~Antiga, L.~Botti, A.~Remuzzi, pyns: An open-source framework for
  0d haemodynamic modelling, Annals of biomedical engineering (2014) 1--13.

\bibitem{Suter:2015}
J.~L. Suter, D.~Groen, P.~V. Coveney, Chemically specific multiscale modeling
  of clay–polymer nanocomposites reveals intercalation dynamics, tactoid
  self-assembly and emergent materials properties, Advanced Materials 27~(6)
  (2015) 966--984.

\bibitem{Suter:2015-2}
J.~L. Suter, D.~Groen, P.~V. Coveney, Mechanism of exfoliation and prediction
  of materials properties of clay–polymer nanocomposites from multiscale
  modeling, Nano Letters\href {http://dx.doi.org/10.1021/acs.nanolett.5b03547}
  {\path{doi:10.1021/acs.nanolett.5b03547}}.

\bibitem{Ray:2014}
S.~S. Ray, Recent trends and future outlooks in the field of clay-containing
  polymer nanocomposites, Macromol. Chem. Phys. 215~(12) (2014) 1162--1179.

\bibitem{Wan:2015}
S.~Wan, B.~Knapp, D.~W. Wright, C.~M. Deane, P.~V. Coveney, Rapid, precise, and
  reproducible prediction of {Peptide-MHC} binding affinities from molecular
  dynamics that correlate well with experiment, Journal of Chemical Theory and
  Computation 11~(7) (2015) 3346--3356.

\bibitem{Sadiq:2008-2}
S.~K. Sadiq, D.~Wright, S.~J. Watson, S.~J. Zasada, I.~Stoica, P.~V. Coveney,
  Automated molecular simulation based binding affinity calculator for
  ligand-bound hiv-1 proteases, Journal of chemical information and modeling
  48~(9) (2008) 1909--1919.

\bibitem{Sadiq:2010}
S.~K. Sadiq, D.~W. Wright, O.~A. Kenway, P.~V. Coveney, Accurate ensemble
  molecular dynamics binding free energy ranking of multidrug-resistant hiv-1
  proteases, Journal of Chemical Information and Modeling 50~(5) (2010)
  890--905.

\bibitem{Pearlman:1995}
D.~A. Pearlman, D.~A. Case, J.~W. Caldwell, W.~S. Ross, T.~E. Cheatham,
  S.~DeBolt, D.~Ferguson, G.~Seibel, P.~Kollman, Amber, a package of computer
  programs for applying molecular mechanics, normal mode analysis, molecular
  dynamics and free energy calculations to simulate the structural and
  energetic properties of molecules, Computer Physics Communications 91~(1)
  (1995) 1--41.

\bibitem{Kollman:2000}
P.~A. Kollman, I.~Massova, C.~Reyes, B.~Kuhn, S.~Huo, L.~Chong, M.~Lee, T.~Lee,
  Y.~Duan, W.~Wang, O.~Donini, P.~Cieplak, J.~Srinivasan, D.~A. Case, T.~E.
  Cheatham, Calculating structures and free energies of complex molecules: 
  combining molecular mechanics and continuum models, Accounts of Chemical
  Research 33~(12) (2000) 889--897.

\bibitem{Case:2005}
D.~A. Case, T.~E. Cheatham, T.~Darden, H.~Gohlke, R.~Luo, K.~M. Merz,
  A.~Onufriev, C.~Simmerling, B.~Wang, R.~J. Woods, The amber biomolecular
  simulation programs, Journal of computational chemistry 26~(16) (2005)
  1668--1688.

\bibitem{FESetup}
{FESetup - http://www.hecbiosim.ac.uk/fesetup/download/0-/3-fesetup} (2015).

\bibitem{Homeyer:2013}
N.~Homeyer, H.~Gohlke, {FEW}: A workflow tool for free energy calculations of
  ligand binding, Journal of Computational Chemistry 34~(11) (2013) 965--973.

\bibitem{Wang:2015}
L.~Wang, Y.~Wu, Y.~Deng, B.~Kim, L.~Pierce, G.~Krilov, D.~Lupyan, S.~Robinson,
  M.~K. Dahlgren, J.~Greenwood, D.~L. Romero, C.~Masse, J.~L. Knight,
  T.~Steinbrecher, T.~Beuming, W.~Damm, E.~Harder, W.~Sherman, M.~Brewer,
  R.~Wester, M.~Murcko, L.~Frye, R.~Farid, T.~Lin, D.~L. Mobley, W.~L.
  Jorgensen, B.~J. Berne, R.~A. Friesner, R.~Abel, Accurate and reliable
  prediction of relative ligand binding potency in prospective drug discovery
  by way of a modern free-energy calculation protocol and force field, Journal
  of the American Chemical Society 137~(7) (2015) 2695--2703.

\bibitem{Pronk:2015}
S.~Pronk, I.~Pouya, M.~Lundborg, G.~Rotskoff, B.~Wesén, P.~M. Kasson,
  E.~Lindahl, Molecular simulation workflows as parallel algorithms: The
  execution engine of copernicus, a distributed high-performance computing
  platform, Journal of Chemical Theory and Computation 11~(6) (2015)
  2600--2608.

\bibitem{Pronk:2013}
S.~Pronk, S.~Pall, R.~Schulz, P.~Larsson, P.~Bjelkmar, R.~Apostolov, M.~R.
  Shirts, J.~C. Smith, P.~M. Kasson, D.~van~der Spoel, B.~Hess, E.~Lindahl,
  {GROMACS} 4.5: a high-throughput and highly parallel open source molecular
  simulation toolkit, Bioinformatics.

\end{thebibliography}







\end{document}